\documentclass[fleqn,10pt]{wlscirep}
\usepackage[utf8]{inputenc}
\usepackage[T1]{fontenc}
\usepackage{comment}
\usepackage{lineno}
\usepackage{braket}

\definecolor{ngreen}{rgb}{0.0,0.5,0.0}

\newcommand{\ams}[1]{{\color{teal}\bf\small[{#1}]}}

\newcommand{\blk}{\color{black}}
\newcommand{\red}{\color{red}}

\title{Experimental evidence that a photon can spend a negative amount of time in an atom cloud}

\author[1,*]{Daniela Angulo}
\author[1]{Kyle Thompson}
\author[1]{Vida-Michelle Nixon}
\author[1]{Andy Jiao}
\author[2]{Howard M. Wiseman}
\author[1]{Aephraim M. Steinberg}
\affil[1]{Department of Physics, and Centre for Quantum Information and Quantum Control,  University of Toronto, 60 St. George Street, Toronto, Ontario, Canada M5S 1A7}
\affil[2]{Centre for Quantum Computation and Communication Technology (Australian Research Council), Centre for Quantum Dynamics, Griffith University, Yuggera Country, Brisbane, Queensland 4111, Australia}

\affil[*]{daniela.angulo@mail.utoronto.ca}

\begin{abstract}
When a pulse of light traverses a material, it incurs a time delay referred to as the group delay. Should the group delay experienced by photons be attributed to the time they spend as atomic excitations? However reasonable this connection may seem, it appears problematic when the frequency of the light is close to the atomic resonance, as the group delay becomes negative in this regime. To address this question, we use the cross-Kerr effect to probe the degree of atomic excitation caused by a resonant transmitted photon, by measuring the phase shift on a separate beam that is weak and off-resonant. Our results, over a range of pulse durations and optical depths, are consistent with the recent theoretical prediction that the 
mean atomic excitation time caused by a transmitted photon (as measured via the time integral of the observed phase shift) equals the group delay experienced by the light. Specifically, we measure mean atomic excitation times 
ranging from $(-0.82\pm 0.31) \tau_0$ for the most narrowband pulse to $(0.54\pm 0.28) \tau_0$ for the most broadband pulse, where $\tau_0$ is the non-post-selected excitation time, given by the scattering (absorption) probability multiplied by the atomic lifetime $\tau_{\rm sp}$. 
These results suggest that negative values taken by times such as the group delay have more physical significance than has generally been appreciated. 

\end{abstract}

\begin{document}

\flushbottom
\maketitle

\thispagestyle{empty}

\section*{Introduction}

One of the foundational scenarios in the field of light-matter interaction is the propagation of light through a dielectric medium. Although this has been extensively studied and, in many ways, thoroughly understood (see, for example, \cite{AllenEberlyMonograph}), it has also sparked controversy in specific areas such as the definition of the speed of a propagating electromagnetic signal \cite{brillouin_wave_1960,milonni2004fast} and the mechanisms governing energy transport in dispersive media \cite{Diener1997,Diener1998,RYC1994}. It is also fundamental to nonlinear optics \cite{hau_light_1999} and to technologies such as quantum memories \cite{Liu2001,Specht2011}. Examining energy transport from a quantum perspective gives rise to another question: How does an individual photon spend its time while propagating through the medium? In the case of light far from resonance, a commonly repeated explanation relies on the uncertainty principle \cite{Grynberg_Aspect_Fabre_2010}. Even when atoms are driven by light with a very large detuning $\delta$, goes the story, they can still experience excitation for a brief period, so long as this is of order $1/\delta$. This extra time spent as a non-propagating excitation would be associated with the group delay (viz. the familiar treatment of slow light in terms of polaritons\cite{Lukin2000}). While this $1/\delta$ dependence captures the behavior of the index of refraction far from resonance, one would probably not expect this association to hold for small detunings, where the group delay is well-known to become negative.

When atoms are illuminated by a pulse of resonant light---which we will refer to as the `signal' beam---they become polarized and have some probability of being found in the excited state at any given time. For a single photon input, the number of excited atoms is represented by the operator $\hat{N}_e$, {\em i.e.}, $\langle \hat{N}_e \rangle(t)$ is the single-atom excitation probability multiplied by the total number of atoms. 
The expectation value $\langle \hat{N}_e \rangle(t)$ is always between zero and one, and can be thought of as a probability. It can be measured using a separate `probe' beam coupled to the atoms through a Kerr nonlinear interaction based on atomic saturation: the instantaneous phase shift on the probe beam is proportional to $\langle \hat{N}_e \rangle(t)$. 

We define the average time that the atoms spend in the excited state ($\tau_0$), or \textit{average atomic excitation time}, as the time integral of the expectation value of the number of atoms in the excited state, $\tau_0 \equiv \int \langle \hat{N}_e \rangle(t) dt$. While this quantity is straightforward to calculate from semiclassical theory as $\tau_0 \equiv P_S / \Gamma$---i.e., the atomic lifetime multiplied by the probability of the photon being scattered--- in this work, we are interested in a different (intrinsically quantum) question: if a single resonant photon incident on a medium {\em is transmitted}, how much time do the atoms spend in the excited state? One might imagine that, unlike scattered photons (mostly absorbed within the first optical decay length), transmitted photons interact with the entire sample and hence cause more integrated excitation; alternatively, one might think that excited atoms are those that lead to loss via spontaneous emission, so that post-selecting on only transmitted photons would result in little or no atomic excitation.
%

Our group previously carried out a measurement of this excitation time, \cite{sinclair_measuring_2022} and found that in a particular parameter regime, a transmitted photon spent nearly as much time as an atomic excitation as the average incident photon. The implication that a large fraction of the excited atoms re-emitted in the forward direction was attributed to coherent forward emission originating from the $\pi$ phase-flip that a broadband pulse's envelope picks up when propagating through an optically thick medium \cite{Crisp1970}. However, such a semiclassical picture cannot make a quantitative prediction as to the phase shift imprinted on our probe {\em when the incident photon is transmitted}, because there is no semiclassical equivalent of that postselection; the probe phase can depend on the intensity of the signal beam but not on whether a particular photon is transmitted or not. 

Prompted by that first experiment, our group therefore employed quantum trajectory theory \cite{dalibard_wave-function_1992,carmichael_open_1993,Wiseman2002} and the weak-value formalism \cite{aharonov_how_1988,Dressel_review_WV} to calculate\cite{thompson2023time} the time that a resonant photon spends as an atomic excitation while being transmitted through a medium. Those calculations predicted that this time is {\em always} equal to the group delay of the transmitted photon, even when that quantity is negative, and that this time depends only on the optical depth of the medium and the spectral content of the pulse. Operationally, this means that the nonlinear phase shift written on our probe by a {\em transmitted} photon should change sign when we move into the negative-delay regime (in our earlier work, the bandwidth and optical depth were such that the net delay was positive).
In the experiment reported here, we test this remarkable theoretical prediction for a wide range of parameters, including regimes in which the group delay is negative.

\section*{Description of the measurement}

\begin{figure}[h!]
\centering
\includegraphics[width=1\linewidth]{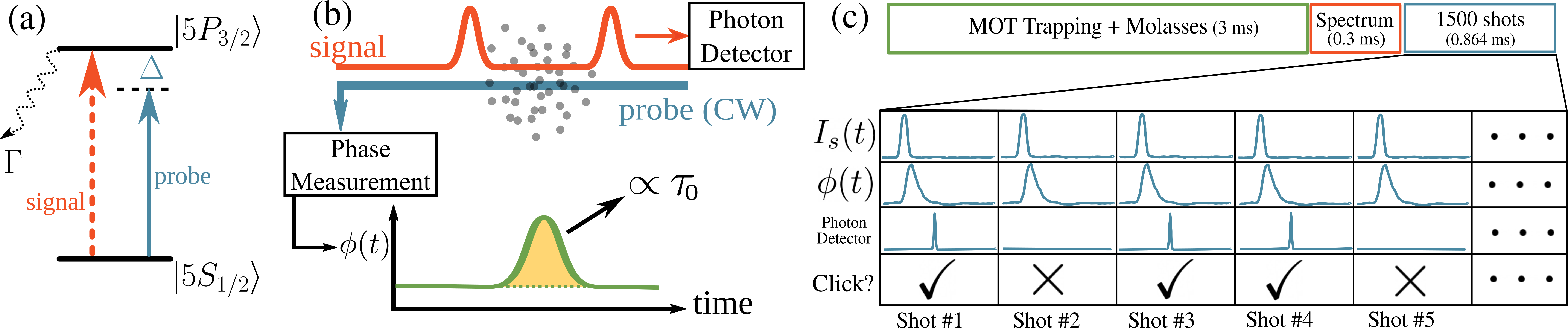}
\caption{Schematics of experimental setup. (a) Atomic level scheme. (b) Conceptual diagram of the experimental apparatus: a resonant pulsed beam (signal) and off-resonant continuous-wave beam (probe) counter-propagate through a cloud of cold $^{85}$Rb atoms, detected at opposite sides of the apparatus. The probe acquires a phase shift $\phi(t)$, proportional to $\langle \hat{N}_e(t) \rangle$, due to the weak saturation of the medium caused by the signal. (c) Timing sequence of each measurement cycle and explanatory chart of data analysis (post-selection).}
\label{fig:setup}
\end{figure}

The Kerr interaction used to probe $\hat{N}_e$ involves two beams of light: a) the `signal' beam, pulsed and resonant with the atomic transition, responsible for causing the excitation time; and b) the probe beam, continuous wave (CW) and off-resonant, which measures the degree of excitation caused by the signal (see Figure \ref{fig:setup}). While the signal pulse propagates through the medium, it weakly saturates the atoms, and
the probe picks up a phase shift proportional to this saturation.

As previously stated, our focus is on the effect of a {\it transmitted} photon. That is, we post-select on the event of transmission. Since our Kerr nonlinearity is very weak and we are interested in the behavior of a system, signal photon, defined by the preparation state and later post-selection, it follows that the quantity we measure can be calculated using the weak-value formalism \cite{aharonov_how_1988}, suitably generalized \cite{thompson2023time}. 
Specifically, the atomic excitation time caused by a transmitted photon, $\tau_T$, is given by the time integral of the real part of the \textit{weak value} of $\hat{N}_e$ \cite{thompson2023time}
, with the initial state specified by the preparation of the input photon and the final state (post-selection) being that of the photon having traversed the medium without scattering. Note that, unlike expectation values, weak values are not restricted to be within the eigenvalue spectrum of the operator being measured, and can even be complex \cite{Josza_complexWV}.

One might expect that, in order to study the behaviour of an individual transmitted photon, we would need to work with single-photon signal pulses at the input. 
However, in our experiment, we leverage the simplicity and rapid rates of coherent-state pulses, combined with the post-selection technique first employed in \cite{Feizpour2015a, Hallaji2017}. In these experiments using coherent states as inputs, our group exploited the fact that upon single-photon detection, the inferred average number of photons present in a medium increases by one; one can thus discern the effect of a single photon on a probe by calculating the difference between that probe's phase in cases with and without a detection event. (This fact was recently proven rigorously, and extended, in\cite{wiseman_obtaining_2023}.) 
This ensures that our measurement of $\tau_T$ using coherent states should match the results of one being performed with single photons. In our experiment, subtracting the probe phase in cases without a detector click $\phi_{NC}(t)$ from cases with a click $\phi_{C}(t)$ allows us to determine the phase shift imparted by a transmitted photon, denoted as $\phi_T(t)=\phi_C(t)-\phi_{NC}(t)$, whose integral is proportional to $\tau_T$. 

The atomic sample is a cold cloud of $^{85}$Rb at a temperature of 60-70 $\mu$K, prepared in a magneto-optical trap formed using two beams of light and a magnetic field gradient of 10 G/cm. The `trap' beam is -25 MHz detuned from the cycling transition of the D2 line ($F=3$ to $F'=4$), while the `repump' beam is resonant with $F=2$ to $F'=3$. The probe and signal counter-propagate, overlapped with each other, and are focused to a waist of 25 $\mu$m inside the atomic cloud of around 1 mm of length. Both the signal and probe address the same atomic transition $F=3$ to $F'=4$ of the D2 line, with a lifetime of $\tau_{\rm sp}\approx26$ ns. The signal is a Gaussian pulse centered on resonance, while the CW probe is detuned by $\Delta\approx-20$ MHz (see Figure \ref{fig:setup}). Additionally, we introduce a sideband at +100 MHz relative to the carrier (probe), with the carrier power ranging from 6-9 nW (about 15\% of resonant saturation intensity) and the sideband being three times stronger. This sideband serves to establish a beat-note interferometer and acts as the phase reference, remaining unaffected by the presence of atoms. We found that setting the probe carrier as far from resonance as the experimental capabilities allowed was the optimal approach for conducting the experiment (see the Supplementary information).

After the interaction inside the atoms has taken place, each beam is collected at opposite sides of the setup. The probe is coupled into a multi-mode fiber and sent to an avalanche photodiode (APD) operating in the linear gain region. The APD's output is subjected to IQ demodulation to extract the probe's phase and amplitude. The signal beam is optically attenuated to ensure that the subsequent power coupled into a single-mode fiber and transmitted to a single-photon counting module (SPCM) prevents multiphoton counts. Furthermore, to eliminate the possibility of spurious correlations driven by the electronic pulse from the SPCM (TTL), we introduce an optical delay of approximately 480 ns before the SPCM using a 100 m-long optical fiber. We use signal pulses with an incident mean photon number of approximately $|\alpha|^2\approx 100$. The attenuation after the atoms was adjusted to maintain an SPCM firing probability of $\sim 0.2$ during a time window of 576 ns, which we refer to as a `shot' and contains a single signal pulse. The choice of a 576 ns duration for each shot provides sufficient spacing between the pulses, ensuring that the dynamics of each pulse remain independent of the others. 

Each measurement cycle begins with a 3-4 ms period dedicated to trapping and cooling the atoms, during which the magnetic field is active for the first 2 ms. Subsequently, the trap beam is turned off, and to prevent atoms from falling to a dark ground state, the repump beam remains active during data acquisition, which begins immediately. Samples are captured every 16 ns. 

Each trace of data starts with two frequency scans of the probe: one with atoms present and the other without atoms. These scans, lasting 300 µs each, serve to calibrate the resonant optical depth (OD) of the medium, which matches the one experienced by the signal. We then send a sequence of 1500 signal pulses, one every 576 ns, totalling 864 µs of data acquisition, for a duty cycle of $\sim25$\%. During this time (see Figure \ref{fig:setup}), we record both the phase and amplitude of the probe, in addition to monitoring the TTL signal from the SPCM.

We anticipated that the phase shift imparted by a transmitted photon would be on the same order of magnitude as the phase shift imparted by the average photon (without post-selection) $\phi_0(t)$, whose peak value ranges from 10 to 20 $\mu$rad. This experimental value is consistent with an estimate given by the inverse of the number of atomic cross-sections addressed by our beams, which is approximately $10^4$. Detecting such a minute effect presents significant challenges. Our phase noise is $\sim$120 mrad, which is four times greater than the quantum noise expected from 5 nW of probe power. This excess noise stems from the noise equivalent power of our detector, which is 100 fW/$\sqrt{\mathrm{Hz}}$, and a measurement bandwidth of 25 MHz. As a result, our measurement required tens of millions of atom cycles for a single set of parameters, translating to about ten hours of data collection.

\section*{Observations and analysis}
\begin{figure}[h!]
\includegraphics[width=1\linewidth]{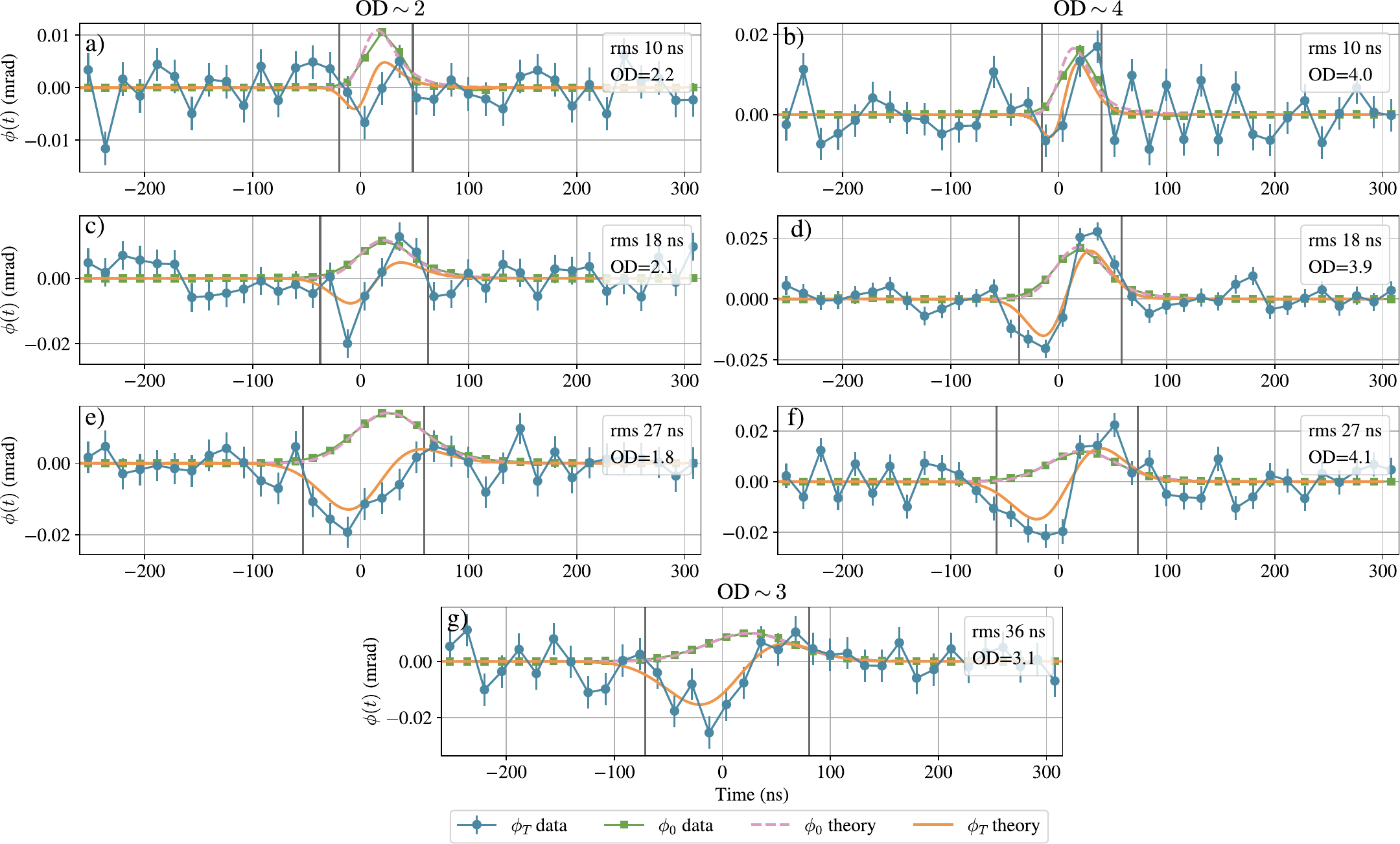}
\caption{Phase shift in a shot (time window containing one pulse) acquired by the probe over time. Blue circles represent the phase shift due to a transmitted photon, with the orange solid line indicating the theoretical expectation. Green squares show the phase shift due to an average incident photon, with the pink dashed line representing the theory. The black vertical lines delimit the integration regions. Error bars depict the standard error of the mean.}
\label{fig:phiTvst}
\end{figure}

We investigated resonant optical depths (OD) of around 2 and 4 (the legends in Figure \ref{fig:phiTvst} contain the exact values), for three pulses with root mean square (rms) durations of 10, 18, and 27 ns. We also took data for a 36 ns rms duration, but the highest OD we could use was 3 due to the increasing presence of background light becoming appreciable for more narrowband pulses (see also the Supplementary information). These choices were made to effectively explore the parameter space while adhering to the technical constraints of our apparatus. 

After synchronizing the measurement of the probe's phase with the corresponding signal detection or lack thereof, accounting for the aforementioned optical delay, we average the phase shift for the shots where the SPCM fired ($\phi_C(t)$) and where it did not ($\phi_{NC}(t)$) and then subtract these two averages. A shot is tagged as `click' if the SPCM fired anywhere in the 576 ns time window. This procedure is repeated for all atom cycles, resulting in a trace of 36 datapoints with $\phi_T(t)$ and the corresponding covariance matrix. The results are depicted in Figure \ref{fig:phiTvst}. The blue dots and green squares represent the data for $\phi_T(t)$ and $\phi_0(t)$, respectively, with error bars for $\phi_T(t)$ indicating the standard error of the mean. The orange solid line and the dashed pink line correspond to the theoretical results for the post-selected and average cross-phase shifts, respectively. These predictions were calculated using the theory in reference \cite{thompson2023time}. These calculations only require the rms duration of the pulse and the average resonant OD seen by the signal during the data runs as inputs.

The theory curves, originally in arbitrary units and time axis, are height-scaled and time-shifted to facilitate comparison with the experimental results (See the Methods section for a full description of the time shifting). The scaling factor, applied to both theory results for $\phi_0(t)$ and $\phi_T(t)$, is chosen such that the peak of the theory for $\phi_0(t)$ matches the peak of the experimental $\phi_0(t)$. The theory presented in Figure \ref{fig:phiTvst} has been adjusted accordingly, using a different scaling factor for each set of parameters while using the same time-shift for all of them. We observe that, for most parameters, the relative size of $\phi_T(t)$ compared to $\phi_0(t)$ measured in our experiment is qualitatively consistent with the theoretical predictions. However, for the case of 18 ns and OD$\sim$2 (see Figure \ref{fig:phiTvst}c), it is evident that the amplitude of the feature exceeds the predicted value.

All of our data is subjected to low-pass filtering at 25 MHz to reduce noise in the phase measurement. This filtering smooths the sharp features, especially noticeable in the 10 ns data, mildly altering the original shape. Despite the fact that the theory curves are not filtered, we still observe good qualitative agreement between the experimental data and the theoretical curves in the horizontal direction. This indicates that, within the error bars, the peaks and troughs of the data and theory coincide in their timing.

\begin{figure}[h!]
\centering
\includegraphics[width=0.7\linewidth]{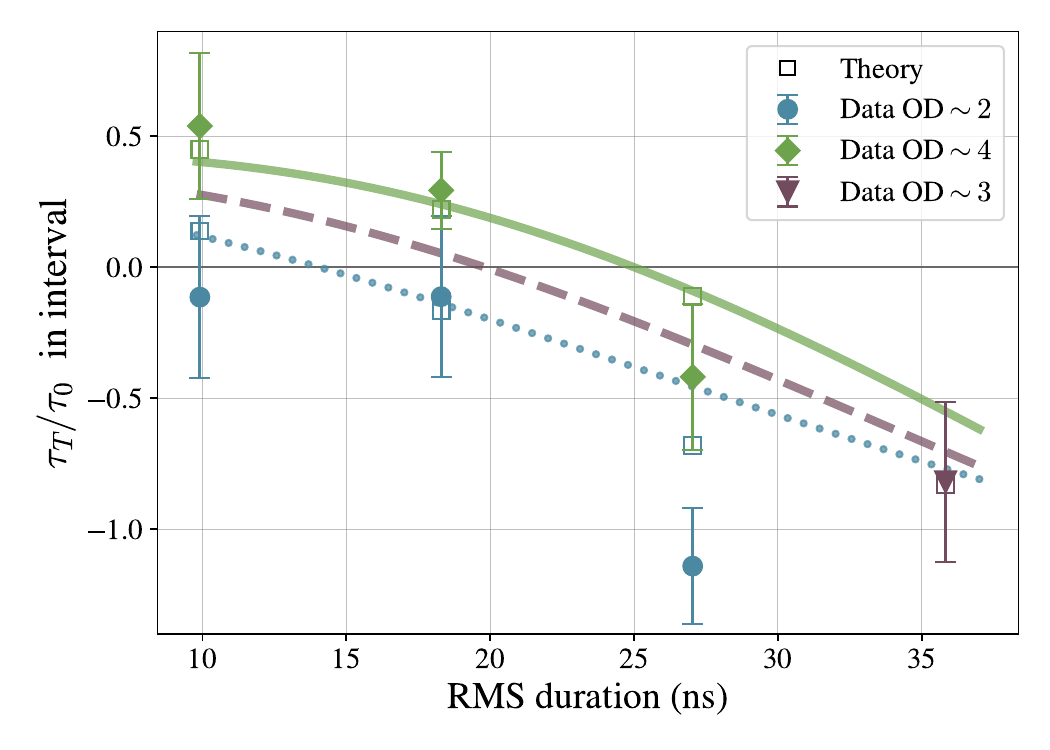}
\caption{Atomic excitation times depicted as the ratio $\tau_T/\tau_0$, obtained through integration over the regions specified in Figure \ref{fig:phiTvst}. Blue circles represent data at OD$\sim 2$, green crosses represent data at OD$\sim 4$, and the purple triangle represents data at OD$\sim 3$ (exact values of the OD for each point are given in \ref{fig:phiTvst}). The theoretical expectation is shown by hollowed squares. The three curves (dotted, dashed, and solid) illustrate theoretical predictions integrating over all times for OD=$\{2,3,4\}$.  Error bars (one standard deviation) were found using the covariance matrix to account for correlations in $\phi_T(t)$ as outlined in Methods.}
\label{fig:summary}
\end{figure}

We proceed to integrate the traces for $\phi_T(t)$ and $\phi_0(t)$ to determine the ratio of the atomic excitation times due to a transmitted photon ($\tau_T$) and an average photon ($\tau_0$) (see Figure \ref{fig:summary}). 
Within each 576 ns shot, the effect of the signal pulse is confined to the central 100--150 ns and the rest of the data is used to determine the background. We use the theory trace $|\phi_T(t)|$ to select a range of integration which keeps the statistical error bars low without introducing a significant systematic error by discarding too much of the dynamics -- to ensure consistency, we implement this by using the arbitrary cutoff at the point where the
tails of the curve reach 30\% of the peak value. 
The black solid vertical lines in Fig \ref{fig:phiTvst} depict the boundaries of the integration intervals. 

Figure \ref{fig:summary} shows the ratio of $\tau_T/\tau_0$ for four different pulse durations and two ODs distinguished by their respective markers (except for 36 ns, where only one data point was taken). These results demonstrate a robust overall agreement between theory and experiment. However, there are specific cases that merit discussion. For instance, at 27 ns, although Figures \ref{fig:phiTvst}e and \ref{fig:phiTvst}f show good agreement for $\phi_T(t)$,$\tau_T$ exhibits a significantly more negative value than expected, and neither of the two data points aligns with their respective theoretical predictions. In contrast, the data obtained at 18 ns and OD$\sim2$ is consistent with the theoretical value, despite a noticeable discrepancy in $\phi_T(t)$, as shown in Figure \ref{fig:phiTvst}c. 

The main sources of discrepancies between the theory predictions and experimental results are spurious correlations between the probe phase and the signal transmission. In other words, if the frequency of the probe is changing rapidly enough with respect to the repetition rate of the pulses, and in a way that is correlated with the intensity of the signal, one can mistake that fluctuation for the effect of a transmitted photon. Given the sensitivity of this measurement, we conducted several systematic checks for these types of correlations, as detailed in the Supplementary information. 

\section*{Discussion}
The measurement conducted by Sinclair et al. \cite{sinclair_measuring_2022} was limited to one set of parameters due to several sources of noise. To explore a broader range of parameters beyond this previous measurement, we made significant progress in mitigating sources of spurious correlations, such as addressing the presence of ground loops. Additionally, we determined that detuning the probe beam further from the atomic resonance---specifically three times further than in the previous experiment---provides a cleaner background. Sinclair's experiment, using a pulse of 10 ns rms duration and an OD of 4, yielded a ratio of $\tau_T/\tau_0=0.77\pm 0.16$. In our study, we obtained a ratio of $\tau_T/\tau_0=0.54\pm 0.28$ for the same parameters, which is in agreement with the theoretical value of $\tau_T/\tau_0=0.45$ from\cite{thompson2023time}. It is notable that our approach to obtain $\tau_T/\tau_0$ also differs from Sinclair's in that we integrated over a region with multiple data points, rather than relying solely on the peak value. However, this resulted in greater statistical error in our measurement; despite us collecting the same amount of data, $\phi_0(t)$ is a much narrower feature than the integration window. This decision to integrate over a broad time window was informed by insights from theoretical work, knowing that the shape of $\phi_T(t)$ can significantly differ from $\phi_0(t)$. Despite the larger error bar, our result is more reliable as it considers more data that provides valuable information.

The results depicted in Figure \ref{fig:summary} undeniably raise the question: what does it mean for an atom to be excited for a negative time, or, equivalently, for a photon to spend a negative amount of time as an atomic excitation? 
The theory work in \cite{thompson2023time} shows that the atomic excitation time due to a transmitted photon is equal to the group delay given a single photon input. Additionally, it provides an illustrative explanation of how this time can be negative, discussing a similar system in which an anomalous dwell time arises from quantum interference in a cavity.

It turns out that it is also possible to leave aside calculations of the weak value of ``excitation time'' observables entirely, and make a direct prediction for the correlation between the probe phase and the number of transmitted signal photons (which is what we measure experimentally), at least in the limit of single-mode signal and probe beams.
The state of the signal and probe after the interaction has taken place is given by 
\begin{equation}
    \ket{\psi}=\sum |c_{mn}|e^{i\varphi_{mn}}\ket{m}_{p}\ket{n}_{s}.
    \label{eq:state}
\end{equation}
Here, the subscripts $p$ and $s$ denote the probe and signal modes, respectively.

By definition, $\phi_T$ represents the change in the optical phase of the transmitted probe $\phi_{PT}$ per transmitted signal photon $n$. Using equation \ref{eq:state}, it can be expressed heuristically as 
\begin{equation}
    \phi_T\equiv \frac{\partial \phi_{PT}}{\partial n}=\frac{\partial (\partial\varphi_{mn}/\partial m)}{\partial n}.
\end{equation} 
where we have used the fact that the optical phase of the probe is related to the variation of the phase $\varphi_{mn}$ of the state with photon number in the probe mode, which we treat as a partial derivative.
The symmetry of second derivatives allows to interchange the order of partial derivatives, yielding
\begin{equation}
    \phi_T=\frac{\partial (\partial\varphi_{mn}/\partial n)}{\partial m}=\frac{\partial \phi_{ST}}{\partial m}\label{eq:symmetry}.
\end{equation}
This demonstrates that a phase shift is also induced on the transmitted signal, denoted as $\phi_{ST}$, and its change with respect to the number of transmitted probe photons $m$ corresponds to $\phi_T$. The origin of this phase shift on the signal is the AC Stark shift of the atomic levels caused by the far-detuned probe. 

The latter term in Equation \ref{eq:symmetry} can be further expressed, using the chain rule, as 
\begin{equation}
    \frac{\partial \phi_{ST}}{\partial m}=\frac{\partial \phi_{ST}}{\partial \omega_0}\frac{\partial \omega_0}{\partial m}=-\tau_g\frac{\partial \omega_0}{\partial m}.
\end{equation}
Here, $\tau_g$ is the group delay for narrowband light and corresponds to the derivative of the spectral phase acquired by the signal beam with respect to its frequency. 
The previous results can be summarized by
\begin{equation}
    \phi_T=\frac{\partial \phi_{PT}}{\partial n}=-\tau_g\frac{\partial \omega_0}{\partial m}. \label{eq:proportional}
\end{equation}
This indicates that the phase shift on the probe induced by a transmitted photon of the signal $\phi_{T}$, as measured by our experiment, is proportional to the group delay of the signal $\tau_g$ with a scaling factor $\frac{\partial \omega_0}{\partial m}$, which corresponds to the AC Stark shift induced by the probe (refer to the Supplementary information for further details of this calculation). Equation \ref{eq:proportional} shares similarities with the expressions for anomalous drag found in \cite{banerjee_anomalous_2022}, suggesting that the group delay governs other effects in regimes where its meaning may be less apparent.

While it is widely known that the group delay can take on negative values, associated with the peak of the transmitted portion of a pulse appearing at times which may indicate superluminal or negative group velocities, it is commonly argued that this quantity does not correspond to the time anything actually ``spends'' in the medium, but merely to the happenstance of when interference is predominantly constructive.
Our observations, however, show that the group delay is a physically meaningful quantity. It not only provides the location of the centre of a transmitted pulse but also correctly describes the magnitude -- and sign! -- of the effect transmitted photons have on other systems they interact with. 

\bibliography{main}

\section*{Methods}

\subsection*{phase shift calibration}
Before each data run, we conduct a calibration procedure to determine the phase shift imparted by an average input photon ($\phi_0(t)$), specific to the chosen set of parameters (OD and pulse duration). This value serves to assess the overlap between the probe and signal inside the atoms and ensures  consistency in experimental conditions from day to day. The gaussian pulses corresponding to the signal are generated by sending the corresponding waveform to an acousto-optical modulator that pulses the light. We take multiple data points for the peak phase shift of the probe as a function of the number of photons in the signal pulse (while keeping the power below 2000 photons where saturation occurs). Subsequently, we fit the data to a straight line and extract the slope, which corresponds to the peak value of $\phi_0(t)$. 

\subsection*{Data analysis and error bars}
The data analysis begins by synchronizing the phase of the probe with the corresponding signal detection or lack thereof, while considering the prior delay of 400 ns introduced to prevent unwanted correlations. For each atom cycle, we obtain a trace for $\phi_T(t)$ by averaging the phase shift $\phi(t)$ for the shots where the SPCM fired ($\phi_C(t)$) and where it did not ($\phi_{NC}(t)$) and subtracting these two. Additionally, we compute the outer product of $\phi_T(t)$ to construct the covariance matrix. We repeat this procedure for all the atom cycles, resulting in a trace with $\phi_T(t)$ and the corresponding covariance matrix, denoted as \textbf{M}. This matrix \textbf{M} contains the information about correlations between any pair of elements of $\phi_T(t)$. The error bars depicted in \ref{fig:phiTvst} correspond to the square root of the diagonal elements of \textbf{M}, representing the standard deviation of the mean. To integrate the traces from \ref{fig:phiTvst} and obtain values for $\tau_T/\tau_0$ shown in \ref{fig:summary}, we use the trapezoidal rule. This integration can be seen as a multivariate function denoted as $f(x_i,...,x_n)$, where $i$ runs over the indices of the interval of integration as specified in \ref{fig:phiTvst} in the main text. We calculate the Jacobian matrix defined as $\mathbf{J_i}=\frac{\partial f}{ \partial x_i}$ and find the error bar of the value of the integral as $\sigma^2=\mathbf{J^T M_R J}$, where $\mathbf{J^T}$ is the transpose of the Jacobian matrix and $\mathbf{M_R}$ is the covariance matrix including only the information related to the points in consideration by the interval of integration.

\subsection*{Data and theory comparison}

The theory curves are output on a time axis determined by the center of the input pulse. However, determining the center of the signal pulse in the experiment presented challenges for us. To align the time axes, we used $\phi_0(t)$ for the most narrowband set of parameters, as for this case, $\phi_0(t)$ closely follows and resembles the shape of the Gaussian input pulse. In our experiment, this occurs at 36 ns (OD $\sim 3$). We obtained the experimental and theoretical values of $\phi_0(t)$ and fitted them to Gaussian profiles. Subsequently, we calculated the time difference between the centers of the fitted Gaussians, which was found to be 252 ns. To match the time axis of the theory, we adjusted the experimental values of both $\phi_0(t)$ and $\phi_T(t)$ by subtracting this time difference from the time variable. Consistently applying this adjustment across all data points, we created the plots shown in Figure \ref{fig:phiTvst}.

The time alignment procedure solely relies on the average phase shifts and assumes a Gaussian lineshape for $\phi_0(t)$ across all pulse lengths, with negligible dispersion from elements other than the atomic medium. The assumption of Gaussianity in $\phi_0(t)$ breaks down when the pulse length goes well below $\tau_{\rm sp}$, which includes, in this case, the data points taken at 10 ns; the shape of $\phi_0(t)$ is closer to an error function for the first half, followed by an exponential decay characterized by $\tau_{\rm sp}$ \cite{sinclair_measuring_2022}.

\end{document}


\title{Supplementary Information: \\ Experimental evidence that a photon can spend a negative amount of time in an atom cloud}

\author{Daniela Angulo}
\affiliation{Department of Physics, and Centre for Quantum Information and Quantum Control,  University of Toronto, 60 St. George Street, Toronto, Ontario, Canada M5S 1A7}
\author{Kyle Thompson}
\affiliation{Department of Physics, and Centre for Quantum Information and Quantum Control,  University of Toronto, 60 St. George Street, Toronto, Ontario, Canada M5S 1A7}
\author{Vida-Michelle Nixon}
\affiliation{Department of Physics, and Centre for Quantum Information and Quantum Control,  University of Toronto, 60 St. George Street, Toronto, Ontario, Canada M5S 1A7}
\author{Andy Jiao}
\affiliation{Department of Physics, and Centre for Quantum Information and Quantum Control,  University of Toronto, 60 St. George Street, Toronto, Ontario, Canada M5S 1A7}
\author{Howard M. Wiseman}
\affiliation{Centre for Quantum Computation and Communication Technology (Australian Research Council), Centre for Quantum Dynamics, Griffith University, Yuggera Country, Brisbane, Queensland 4111, Australia}

\author{Aephraim M.  Steinberg}
\affiliation{Department of Physics, and Centre for Quantum Information and Quantum Control,  University of Toronto, 60 St. George Street, Toronto, Ontario, Canada M5S 1A7}

\date{\today}

\maketitle

\tableofcontents 

\newpage

\section{Technical limitations of the apparatus: background light}

As discussed in the main text, the probe and signal counter-propagate, overlap and are focused inside the atoms. This overlap implies that the collection stage for the probe (signal) is susceptible to collect undesired light from the signal (probe). This light can come from back-reflections off surfaces perpendicular to the direction of propagation of the beams or back-scattered light by the atoms. This effect is more noticeable in the case of the signal detection, given the relative strengths of the two beams and the fact that any trigger of the SPCM within a time window of 576 ns is considered a successful postselection. All the incident signal pulses (10, 18, 27 and 36 ns rms duration) contain a mean photon number of $|\alpha|^2\approx 100$ and are further attenuated after passing through the atoms depending on the OD, and the pulse duration. In contrast, the probe input has approximately 300 photons per 16 ns bin, resulting in 50 times more probe photons in a shot. This increases the likelihood of probe light getting into the signal collection fiber; for our experiment, around 90\% of the spurious counts come from probe light back-scattered from the atoms. 

The presence of light from other sources in the signal collection is detrimental to our measurement because detection events from a different light source, other than the signal, can also exhibit correlations with the phase of the probe. These correlations can be mistaken for the effect of a transmitted signal photon. 

The ratio between probe (spurious) and signal clicks at the SPCM worsens as the pulse duration and OD increase i.e., more OD implies more signal absorption while increasing the amount back-scattered probe light from the atoms. We decided to operate below a 0.1 ratio, which imposed limitations on the parameter space we could explore.

\section{Theoretical connection between nonlinear phase-shift and group delay}

In the main text, Equation 6 establishes a proportionality relation between the phase shift imparted by a transmitted photon and the group delay. Now, we aim to demonstrate that, for the conducted experiment, the measured quantity defined as the atomic excitation time caused by a transmitted photon, denoted as $\tau_T$, corresponds to the group delay, denoted as $\tau_g$. 

As previously discussed, the atomic excitation time caused by a photon, represented as the time integral of the number of atoms excited by the signal, $N_e^s(t)$, is proportional to the integral of the phase shift imparted on the probe, $\phi(t)$. The relationship between these two quantities, in the limit of weak probe and weak signal, is approximately 
\begin{equation}
    \int\phi(t)dt=\frac{2}{\Gamma}\frac{\Delta}{1+(2\Delta/\Gamma)^2}\frac{\sigma_0}{A}\int N_e^s(t) dt,
\end{equation}
where A is the beam area, $\sigma_0$ is the atomic cross-section on resonance, $\Delta$ is the probe detuning, and $\Gamma$ is the decay rate of the atomic transition. We denote the multiplicative factor in front of the integral of the number of excited atoms as $C$. Thus, we can simply state that 
\begin{equation}
    \int\phi_T(t)dt=C\tau_T. \label{eq:conversion}
\end{equation} Here, $\tau_T$ represents the integral in time of the number of atoms excited exclusively by transmitted photons. From this, we can identify $C$ as the conversion factor between phase-shift and atomic excitation time.

Recall that the proportionality constant $\frac{\partial \omega_0}{\partial m}$ in Equation 6 in the main text can be seen as the frequency change in the atomic resonance, due to AC stark-shift, per probe photon. The AC Stark shift of the atomic level due to the probe beam is given by 
\begin{equation} \label{eq:AC_Stark}
    \delta \omega_{AC}=-\frac{2\Omega^2}{\Gamma^2}\frac{\Delta}{1+(2\Delta/\Gamma)^2}=-\frac{I}{I_{sat}}\frac{\Delta}{1+(2\Delta/\Gamma)^2},
\end{equation}
where $I$ is the beam intensity, $I_{sat}$ is the saturation intensity and $\Omega$ is the Rabi frequency. 

There are $N_p$ photons on average within a range of integration of duration T . 
Replacing Equation \ref{eq:AC_Stark} in Equation 6 from the main text and taking the integral of both sides gives
\begin{equation}
\begin{split}
\int \phi_T dt & =-\tau_g\int \frac{\delta \omega_{AC}}{N_p} dt.\\
&=\frac{\hbar\omega_p}{AI_{sat}}\frac{\Delta}{1+(2\Delta/\Gamma)^2}\tau_g\\
&=\frac{\sigma_0}{A\Gamma}\frac{\omega_p}{\omega_0}\frac{\Delta}{1+(2\Delta/\Gamma)^2}\tau_g\\
&=C \frac{\omega_p}{\omega_0} \tau_g,
\end{split}\label{eq:split}
\end{equation} 
where we have used the fact that the resonant atomic cross-section is $\sigma_0=\frac{\hbar\omega_0\Gamma}{2I_{sat}}$. Using Equation \ref{eq:conversion} combined with the result from Equation \ref{eq:split} gives
\begin{equation}
    \tau_T=\frac{\omega_p}{\omega_0} \tau_g\approx \tau_g,
\end{equation}
due to the proximity of the absolute frequencies $\omega_p$ and $\omega_0$, both on the order of 400 THz and detuned from each other by only 20MHz.

\section{Checking for systematic effects}
In our experiment, we measure the nonlinear phase-shift of the probe imparted by a transmitted photon $\phi_T(t)$. However, this phase-shift is expected to be six orders of magnitude smaller than the linear phase acquired by the probe beam due to the atoms alone, in absence of the signal. This significant difference in magnitude makes it easy for any parasitic effect causing phase fluctuations in the probe correlated with the signal intensity to potentially obscure or overwhelm such a small $\phi_T(t)$. Given that the post-selection is done in a time window of 576 ns, the fluctuations must be faster than $\sim 1$ MHz in order to disturb our measurement significantly. 

We addressed several sources of spurious correlations and conducted checks for systematic effects. A null result in these tests indicates that the integral of $\phi_T(t)$ in the region of interest is consistent with zero. The region of integration is established following the procedure outlined in the main text. While a null result does not guarantee the accuracy of the $\phi_T(t)$ measurement, it ensures that the simultaneous effects of signal, probe and atoms are considered when taking any data of $\phi_T(t)$.

\subsection{No atoms}
This test involves sending the probe and the signal through their usual paths (counterpropagating through the vacuum chamber) without encountering any atoms. To ensure no atoms are trapped in our MOT, we block the trap and repump light and turn off the magnetic field. 

\begin{figure}[h!]
\centering
\includegraphics[width=1\linewidth]{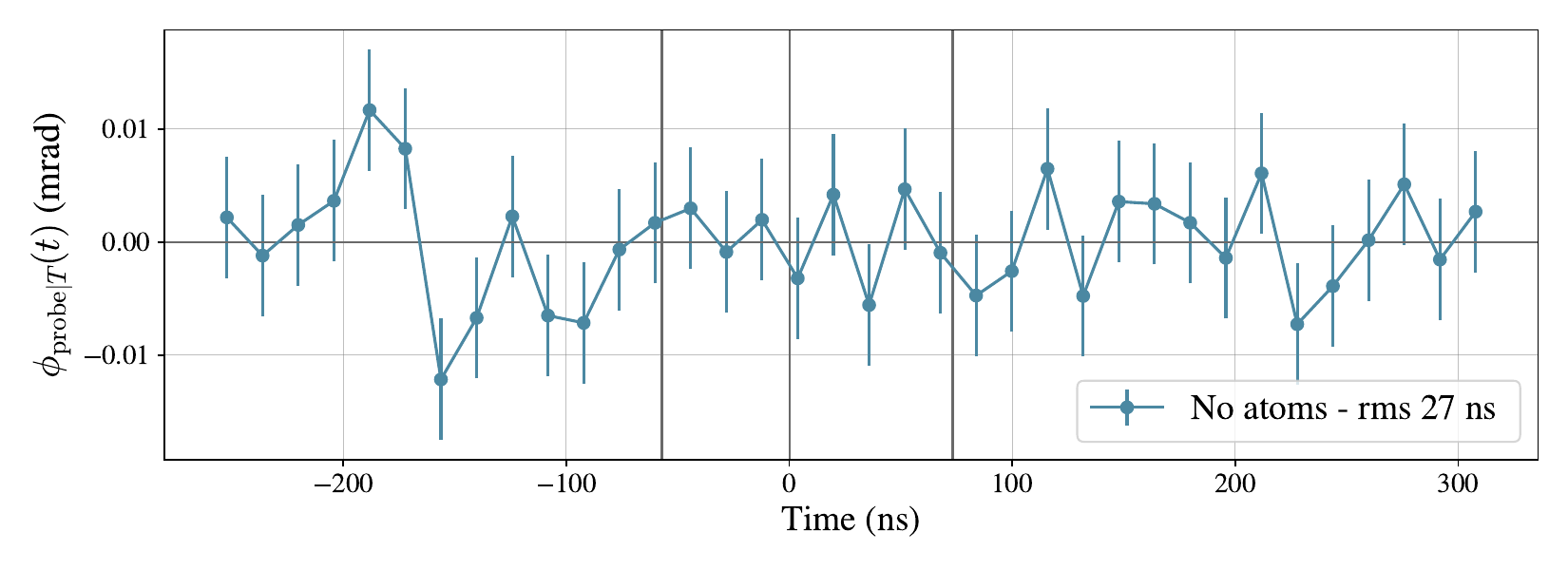}
\caption{Systematic check for correlations between the phase of the probe and the probability of detecting signal light when no atoms are present in the chamber. Integration window is depicted in black horizontal lines.}
\label{fig:noatoms}
\end{figure}

A null result here shows the absence of electrical correlations between the phase of the probe and the intensity of the signal, meaning any correlations present would be mediated by the atomic medium. Figure \ref{fig:noatoms} shows $\phi_T(t)$ as a function of time using pulses of 27 ns rms duration. The integral gives a ratio of $\tau_T/\tau_0=0.07\pm0.29$ giving no indication of a parasitic effect. In order to normalize to $\tau_0$ and define the integration region we used the $\phi_0(t)$ data for an OD of 4 and pulses of 27 ns rms. 

\subsection{Signal bypassing atoms}

When initially conducting this test, we found broad features in $\phi_T(t)$ with amplitudes approximately ten times larger than $\phi_0(t)$, around 50$\mu$rad. This phenomenon stemmed from rapid changes in the probe frequency (at a frequency of 2 MHz), which were correlated with the signal intensity controlled by an acousto-optic modulator. After months investigation, we worked hard to isolate different electrical circuits and remove such correlations.

To confirm the absence of these kind of correlations, we performed a test by sending the signal light directly to the SPCM without passing it through the atomic sample. While the atoms and the probe light interacted as usual, the signal light was merely collected. We used the standard settings to generate signal pulses of 36 ns rms duration containing 100 photons on average, but we significantly increased their attenuation to compensate for the absence of the atoms. This measurement tests the presence of electrical correlations between the probe frequency and signal intensity.

\begin{figure}[h!]
\centering
\includegraphics[width=1\linewidth]{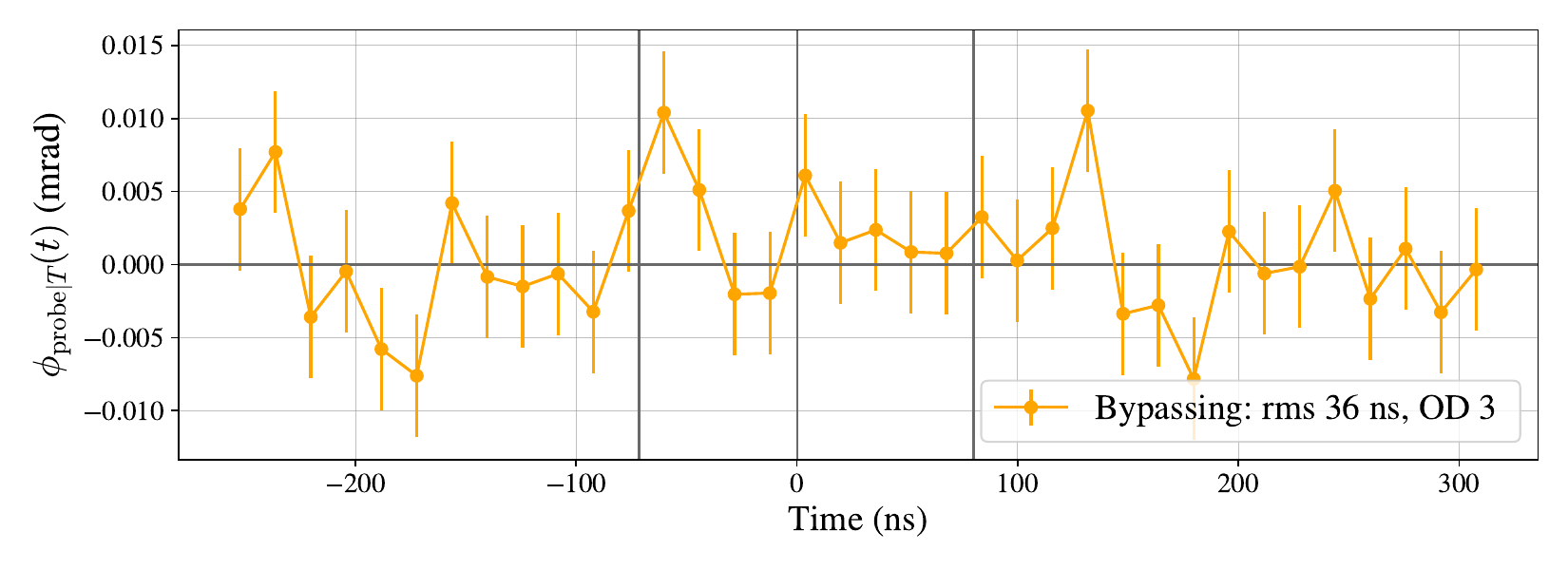}
\caption{Systematic check for correlations between the phase of the probe and the probability of detecting signal light when bypassing the atoms with the signal. Integration window is depicted in black horizontal lines.}
\label{fig:bypassing}
\end{figure}

Figure \ref{fig:bypassing} shows the data for this test obtained using pulses of 36 ns rms duration and an OD of 3. The resulting ratio in the integration region, delimited by black lines, is $\tau_T/\tau_0=0.43\pm0.23$. We take this as a rough estimate of the size of the systematic error. However, it is noteworthy that the plotted data does not closely resemble the time dynamics displayed by the data presented in the main text or predicted by the theory.

\subsection{No signal beam}
\begin{figure}[h!]
\centering
\includegraphics[width=1\linewidth]{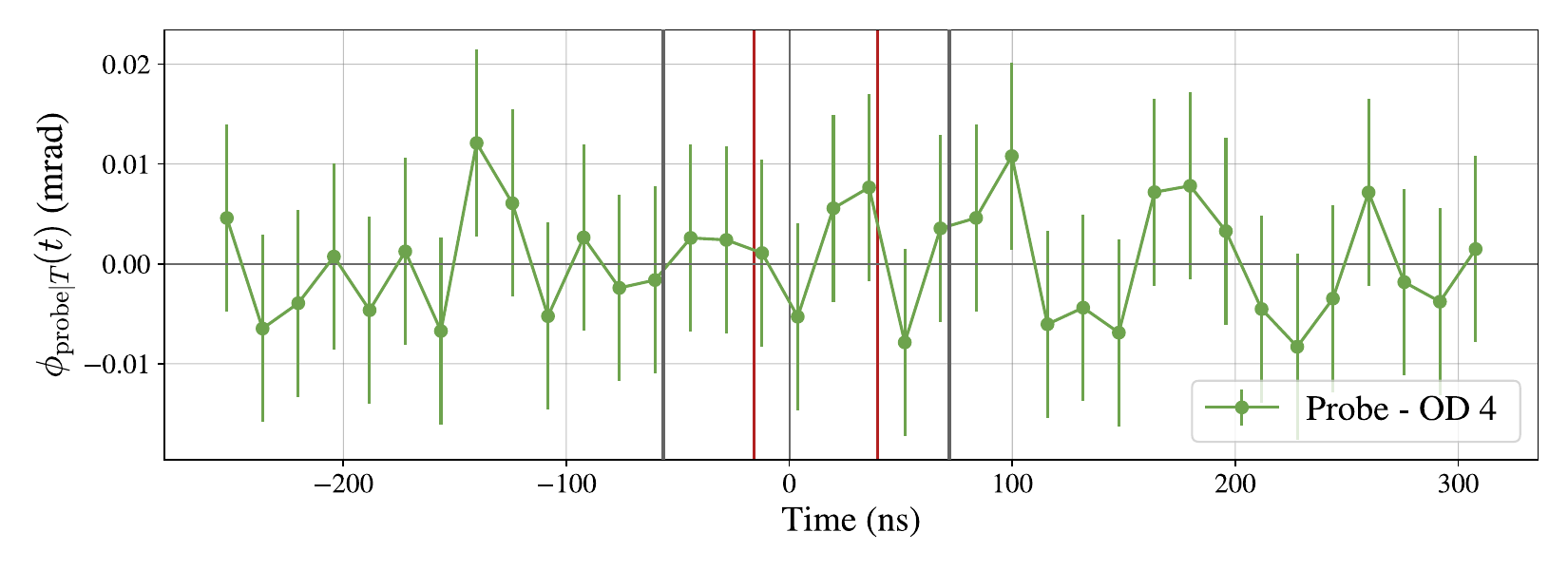}
\caption{Systematic check for correlations between the phase of the probe and the probability of detecting reflected probe light with the signal beam blocked. Integration windows are depicted in red and black horizontal lines.}
\label{fig:probe}
\end{figure}

After addressing the noise source responsible for correlations in the previous test, we reintroduced the atoms and attempted to measure $\phi_T(t)$ with all components in place. However, we still observed features of a similar magnitude as before (approximately 50 $\mu$rad). After extensive investigation, we identified the laser controller of the probe as the culprit, introducing significant and rapid (2 MHz) frequency fluctuations that correlated the SPCM clicks and the phase of the probe. Recall that due to the geometry of the experiment, some probe light gets coupled in the signal collection. The correlations induced by this noise would have to give $\phi_T(t)$ features of milliradian size to persist in our measurements, despite the probe clicks only accounting for as much as 10\% of the total clicks. We confirmed this by measuring $\phi_T(t)$ while blocking the signal beam and removing all attenuation from the signal collection. Fortunately, the problem was easily resolved by switching to a different laser controller. The data in Figure \ref{fig:probe} was taken after fixing the problem using an OD of 4. It gives $\tau_T/\tau_0=0.19\pm0.62$ for the narrow integration window (depicted with red lines) and $\tau_T/\tau_0=0.16\pm0.50$ for the longest integration window used (depicted with black lines). Accounting for the fact the probe clicks could only be as much as 10\% of the total clicks, these two results would contribute at most $\tau_T/\tau_0=0.02\pm0.06$ and $\tau_T/\tau_0=0.02\pm0.05$ of systematic error to the measurement of $\tau_T/\tau_0$.

\section{Optimal probe detuning}
Our first approach to determining the optimal probe detuning for conducting the experiment involved optimizing the signal-to-noise ratio (SNR) of the phase measurement. This optimization relied on two main criteria. Firstly, we evaluated the amplitude of the phase shift imparted on the probe per signal photon $\phi_0$ as a function of probe detuning, which ideally follows a dispersive Lorentzian. Secondly, we aimed to mitigate unwanted correlations by identifying a detuning at which frequency fluctuations do not cause a significant change in phase shift, i.e., where the derivative of the linear phase of the probe with respect to frequency is small. Both of these requirements were satisfied by placing the probe at $\pm \Gamma/2$.

Initially, using $\pm \Gamma/2$ appeared to be the best choice. However, when measuring the amplitude of the XPS as a function of probe detuning, we consistently observed that the phase-shift was approximately two times larger for negative detunings compared to positive ones. This phenomenon was tentatively attributed to a `dipole-trap' effect, wherein the atoms were pushed (pulled) due to the positive (negative) detuning of the probe light. Consequently, we opted to operate at negative detunings. 

We also explored the change in the optical depth (OD) experienced by the probe within an atom cycle (900$\mu$s); it decreases from an initial value as time advances. We observed that the percent change was lower as we moved further from resonance. Based on these findings, we opted to operate at -10 MHz, which is further from resonance compared to the previous experiment conducted at -5 MHz. This new detuning ensured that the OD, within an atom cycle, would not decrease by more than 10\% of its initial value. 

\begin{figure}[h!]
\centering
\includegraphics[width=1\linewidth]{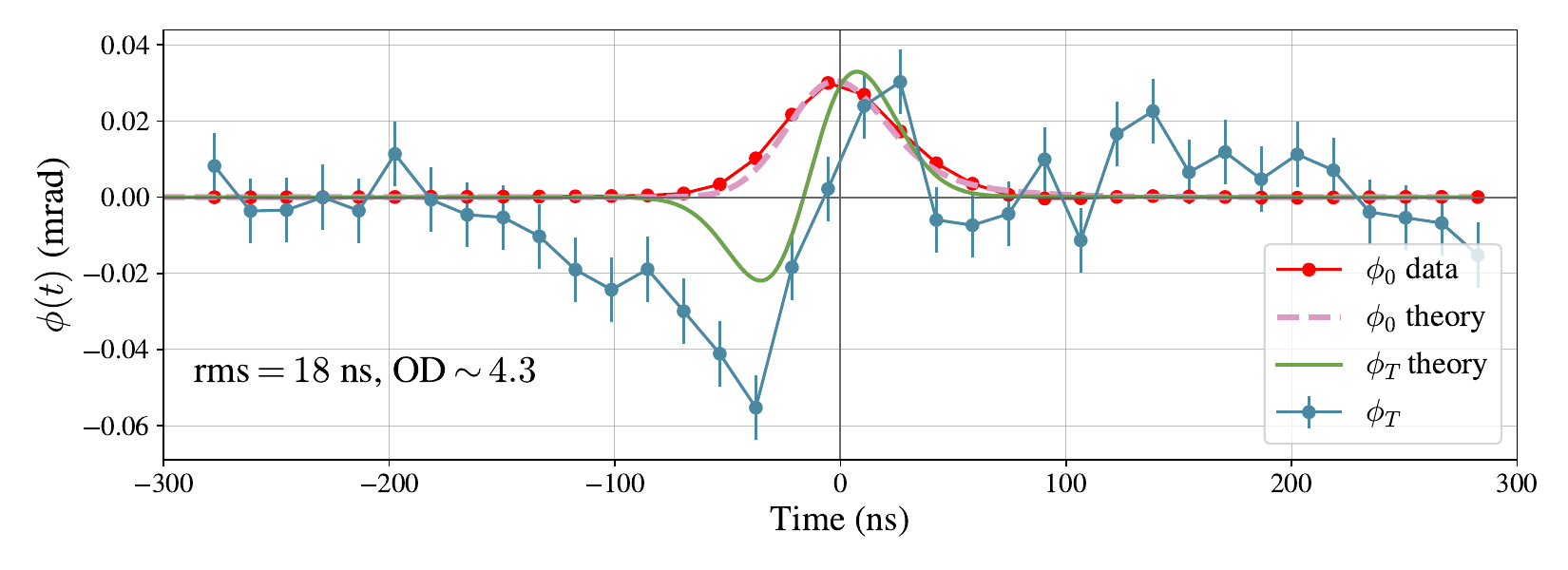}
\caption{Data taken at -10MHz for signal pulses of 18 ns rms and $\mathrm{OD}\sim4.3$.}
\label{fig:probe_det}
\end{figure}

For the first few months, we collected data with the probe at -10 MHz. However, we consistently found that $\phi_T(t)$ lasted much longer than either the data for $\phi_0(t)$ or our theoretical predictions for $\phi_T(t)$ (which completely neglect propagation of the probe). An example of this data is presented in Figure \ref{fig:probe_det}.
We don't yet have a clear understanding of why this happened, but we decided to study the behaviour of $\phi_T(t)$ as we moved further from resonance and found that it eliminated this problem. We then chose to operate at -20 MHz, where all the data presented in this paper was taken. 
